\documentclass[fleqn,twoside]{article}
\usepackage{espcrc2}

\usepackage{graphicx}
\usepackage[figuresright]{rotating}

\newcommand{\AmS}{{\protect\the\textfont2
  A\kern-.1667em\lower.5ex\hbox{M}\kern-.125emS}}

\title{Experimental signatures of non-standard pre-BBN cosmologies\thanks{Talk given at the ``Dark Matter Conference", GGI, Feb. 9-11, 2009, Florence, Italy}}

\author{Graciela B. Gelmini\address[UCLA]{Department of Physics and Astronomy,
 UCLA, 475 Portola Plaza,
 Los Angeles, CA 90095, USA}%
        \thanks{This work was supported in part by the US DOE grant DE-FG03-91ER40662 
  Task C.}}
       
\begin{document}

\begin{abstract}
If detected, dark matter particles, such as WIMPs and sterile neutrinos, will be the earliest relics we can study, the first coming from before Big Bang Nucleosynthesis, an epoch from which we have no data so far. Here we discuss how different pre-BBN cosmological models affect the properties of these particles, and how these properties could thus allow to determine the evolution of the Universe before BBN.
\vspace{1pc}
\end{abstract}

\maketitle

\section{Introduction}

 Dark matter (DM) particle candidates, such as  WIMPs, weakly interacting  massive particles  (also sterile neutrinos and axions)  are produced before Big Bang Nucleosynthesis (BBN). This is an epoch from which we have no data.  BBN is the earliest episode (200 s  after the Bang,  $T\simeq 0.8$ MeV) from which we have a trace, the abundance of light elements D, $^4$He and $^7$Li.   
 In order for BBN and all the subsequent  history of the Universe to proceed as usual, it is enough that the earliest and highest temperature during the last radiation dominated period, the so called reheating temperature $T_{RH}$, is larger than 4 MeV~\cite{Hannestad:2004px}.  

The relic density and relic velocity distribution before structure formation of WIMPs (and other DM candidates) depend on the characteristics of the Universe (expansion rate, composition, etc.) before BBN.  If these particles are ever found, they would be the first relics from that epoch that could be studied. Thus we will want to extract as much information about the Universe at the moment these particles decoupled as we can.  
Will we be able to differentiate cosmological parameters from particle physics parameters that determine their relic density and  relic velocity distribution? That is, will we be able  to discriminate between different pre-BBN cosmologies through studying the DM particles? To start with, we need to know how large are the possible  effects of different viable pre-BBN cosmologies  on DM particle properties we could measure.

\section{Standard and non-standard pre-BBN cosmologies}

The argument showing that WIMPs  are good DM candidates is old. The density per comoving volume of non relativistic  particles in equilibrium in the early Universe decreases exponentially with decreasing temperature, due to the Bolzmann factor, until the reactions which change the particle number become ineffective.  
 At this point, when  the annihilation rate  becomes smaller than the Hubble expansion rate, 
  the WIMP number per comoving volume becomes constant. This  moment of chemical decoupling or freeze-out happens later, i.e. for smaller WIMP densities, for larger annihilation cross sections $\sigma$. If there is no subsequent change of entropy in matter plus radiation, the present relic density is $\Omega_\chi^{\rm std} h^2 \simeq 10^{-10} {\rm ~GeV^{-2}}/ {\left< \sigma v \right> } $, which for weak order $\sigma$ gives the right order of magnitude of the DM density (and a temperature  $T_{f.o.} \simeq m_\chi/20$ at freeze-out for a WIMP $\chi$ of mass $m_\chi$).  
    
\begin{figure}
\includegraphics[width=0.40\textwidth]{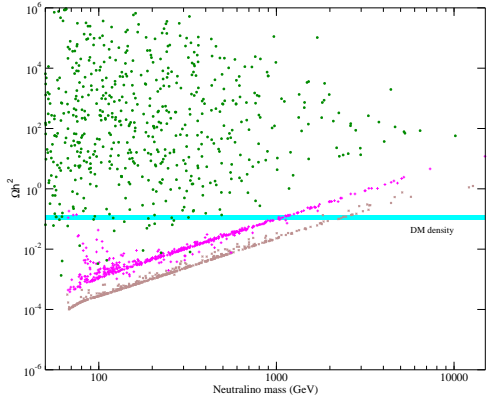}
\vspace{-10pt}
\caption{Standard neutralino relic density $\Omega h^2$ vs mass for 1700 different MSSMs (colo indicate composition: green for bino-like, pink for higgsino-like, brown for  wino-like)~\cite{Gelmini:2006pq}.}
\label{well-tempered}
\vspace{-0.6cm}
\end{figure}

   This is a ballpark argument. When actually applied to particle models, the requirement that the WIMP candidate of the model must have the measured DM density is very constraining. In many supersymmetric models, in which the WIMP candidate is usually a neutralino, this ``DM constraint" 
is very effective in restricting the parameter space of models. With the standard pre-BBN cosmological assumptions,  after the constraints imposed by LEP II, having a neutralino with the DM density
requires special tuned relation among parameters~\cite{ArkaniHamed:2006mb}. In Minimal Supersymmetric Models (MSSMs) that could be found at the LHC, the ``DM constraint" requires a``well tempered neutralino"~\cite{ArkaniHamed:2006mb}, fine  tuned to be at the boundary between a pure bino and a pure higssino or a pure bino and a pure wino.  In fact, as shown in Fig.~\ref{well-tempered}, with the standard cosmological assumptions bino-like neutralinos tend to be overdense (except when they annihilate resonantly or coannihilate with an almost degenerate slepton) while higgsino-like and wino-like are underdense (or with masses close to 1TeV or 2 TeV respectively, beyond the reach of the LHC).  Fig.~\ref{well-tempered} shows the relic density as function of the mass of neutralinos of different composition in 1,700 different MSSMs characterized by nine parameters defined at the electroweak scale~\cite{Gelmini:2006pq}.
In minimal supergravity models (mSUGRA) for instance, the neutralino is typically  bino-like thus its relic density  tends to be larger than observed. The  ``DM constraint"  is found to  be satisfied only along  four very  narrow regions in  the fermionic and scalar mass parameter space $m_{1/2}$, $m_0$ (see e.g. Ref.~1).  Most of the ``benchmark points", special models chosen to study in detail in preparation for the  LHC  and the next possible collider 
(such as A' to L', Snowmass Points and Slopes or SPS 1a',1b, 2, 3 ,4, 5,  Liner Collider Cosmo points or LCC 1,2,3,4) lie on those very narrow bands~\cite{benchmarks}. 
 Neutralinos are underabundant (account for a fraction of the DM) also in narrow regions adjacent to these just mentioned, but  in most of the parameter space neutralinos are overabundant and the corresponding models are many times said to be rejected by cosmology.
Is it  correct to reject all these supersymmetric  models? The answer is  no.
The issue is that the position   in parameter space of the narrow bands just mentioned depends not only on the particle model to be tested in collider experiments, but on the assumptions made about the history of the Universe before BBN. 

The standard computation of relic densities relies  on assuming that radiation domination began before the main epoch of production of the relics and that the entropy of matter and radiation has been conserved during and after this epoch.   In the case of WIMPs it is also assumed that they are produced thermally,  i.e.\ via interactions with the particles in the plasma.
 With these assumptions chemical decoupling happens at $T_{f.o.}\simeq m_{\chi}/20$ and kinetic decoupling, the moment after which WIMPs do not exchange momentum efficiently with the cosmic radiation fluid, happens  later, at $T_{k.d.} \simeq$ 10 MeV-10 GeV~ \cite{Profumo:2006bv}. Thus all WIMPs with $m_\chi \geq 100$ MeV decouple at temperatures higher than 4 MeV, when  the content and expansion history of the Universe may differ from the standard assumptions.

Non-standard pre-BBN cosmological models are more complicated  than standard ones and although several aspects of them have been studied, a comprehensive scenario in each of them is yet missing.
Usually non-standard cosmological  scenarios contain additional parameters that can be adjusted to modify the WIMP relic density. However these are due to physics that does not manifest itself in accelerator or DM detection experiments.

  In non-standard cosmological models, the WIMP relic abundance may be higher or lower than the standard abundance.  The density may  be increased by creating WIMPs from decays of particles or extended objects (non-thermal production) or by increasing the expansion rate of the Universe at the time of freeze-out.
The density may be decreased by  reducing the expansion rate of the Universe at freeze-out,  by reducing the rate of  thermal production (through a low $T_{RH} < T_{f.o.}$) or by producing radiation after freeze-out (entropy dilution).   
 \begin{figure}
\includegraphics[width=0.45\textwidth]{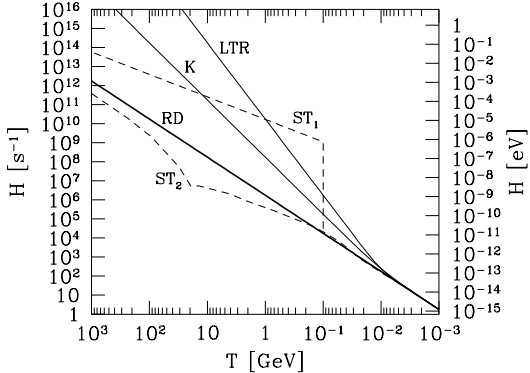}
\vspace{-10pt}
\caption{$H(T)$ for several pre-BBN cosmological models: radiation-dominated (RD),  LTR,
 kination (K) and  scalar tensor with $H$ increase (ST1) and with $H$ decrease (ST2). Fig. from Ref.\cite{GG-Bertone}.}
\label{H(T)}
\vspace{-0.6cm}
\end{figure}
  Any extra contribution to  the energy density of the Universe  would increase the Hubble expansion rate $H$ and lead to larger relic densities (since the decreasing interaction rate becomes smaller than $H$ earlier, when densities are larger). This can happen in   Brans-Dicke-Jordan~\cite{Kamionkowski:1990ni} model, models with anisotropic 
expansion~\cite{Barrow:1982ei,Kamionkowski:1990ni,Profumo:2003hq}, scalar-tensor (ST)~\cite{Santiago:1998ae,Damour:1998ae,Catena:2004ba,Catena:2007ix} or
kination~\cite{Salati:2002md,Profumo:2003hq} models and other 
models~\cite{Barenboim:2006rx}. These models alter  the thermal evolution of the Universe without an extra entropy production. Kination~\cite{Salati:2002md} is a period  in which the kinetic energy $\rho_\phi\simeq  \dot{\phi}^2/2$ of a scalar field $\phi$ (maybe a quintessence field) dominates.
   The relic density of WIMPs which decouple during kination may be increased by a factor $\sqrt{\eta_\phi} 10^{3}  (m_\chi/100 {\rm GeV})$, where  $\eta_\phi = \rho_\phi/ \rho_\gamma< 1$ is the ratio of the $\phi$ and  photon densities  at $T\simeq 1$ MeV. ST models
 of gravity have a scalar field coupled only through the metric tensor to the matter fields. The expansion of the Universe  drives the scalar field towards a state where the theory is indistinguishable from General Relativity at  $T \leq T_\phi$, before BBN.  Fig.~\ref{H(T)} shows $H(T)$ for several  pre-BBN cosmological models.  WIMPs freeze-out before $T_\phi$, when $H$ is larger, and although after $T_\phi$ there is a ``reannihilation phase", the net effect is that WIMPs densities are increased by factors of 10 to 10$^3$~\cite{Catena:2004ba}.
 In  some ST models  $H$ may be decreased, leading to relic densities smaller than standard  by as much as a factor 0.1~\cite{Catena:2007ix}. The factors just mentioned, 10$^3$ to 0.1, are not large enough to bring any MSSM neutralino, for example, to have the DM density. 

 Not only the value of $H$ but the dependence of the temperature $T$ on the scale factor of the Universe $a$ is different  than the usual $a \sim T^{-1}$, if  entropy in matter and radiation is produced. This is the  case if a scalar field $\phi$ oscillating  around its true minimum while decaying is the dominant component of the Universe just before BBN, for which $H \sim T^4 \sim a^{-3/2}$.  Models of this type include some with  moduli fields, either the Polonyi field~\cite{Moroi:1994rs,Kawasaki:1995cy} or others~\cite{Moroi:1999zb}, or an Affleck-Dine field and Q-ball decay~\cite{Fujii:2002kr}, and thermal inflation~\cite{Lyth:1995ka}.
Moduli fields correspond to flat directions in the supersymmetric potential, which are lifted by the same mechanisms that give mass to the supersymmetric particles of the order of  a few to 10's of TeV, and they usually have interactions of gravitational strength.  The decays of the  $\phi$ field  finally reheat the Universe to a low reheating temperature $T_{\rm RH} \simeq 10~{\rm MeV}(m_\phi/\rm
100~TeV)^{3/2}$ ($m_\phi$ is the $\phi$ mass), which could be not much larger than 4 MeV.  In these low temperature reheating (LTR) models there can be direct production of DM relics in  the decay  $\phi$ which  increases the relic density, and there is entropy generation, through the decay of $\phi$ into radiation, which suppresses the relic abundance. The change in relic density in LTR models is larger than in other non standard models.  Both thermal and non thermal production mechanisms in LTR  have been discussed
\cite{McDonald:1989jd,Kamionkowski:1990ni,Moroi:1994rs,Kawasaki:1995cy,Chung:1998rq,Moroi:1999zb,Giudice:2000ex,Allahverdi:2002nb,Khalil:2002mu,Fornengo:2002db,Pallis:2004yy,Gelmini:2006pw,Gelmini:2006pq,Drees:2006vh,Endo:2006ix}, mostly in supersymmetric models where the WIMP is the neutralino. 
  The WIMP relic density depends only on two additional parameters (determined by the model at high energy scales) besides the usual ones:   $T_{\rm RH}$ and $b/m_\phi$, the ratio of $b$, the net number of WIMPs produced on average per $\phi$ decay  and the $\phi$ mass~\cite{Gelmini:2006pw}. $b$ is a highly model dependent parameter~~\cite{Moroi:1999zb,Allahverdi:2002nb,Gelmini:2006pw},
 \begin{figure}
\hspace{-0.3cm}
\includegraphics[width=0.50\textwidth]{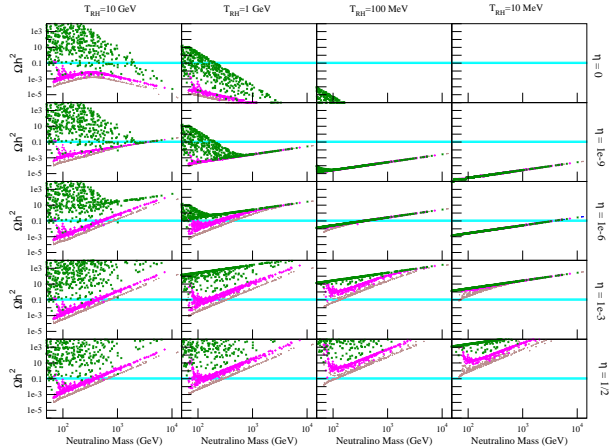}
\vspace{-10pt}
\caption{LRT neutralino relic density $\Omega h^2$ vs mass for different values of $T_{RH}$ and $\eta$ for  the same 1700 MSSMs as in Fig.~\ref{well-tempered}~\cite{Gelmini:2006pq}.}
\label{LTR-omega}
\vspace{-0.6cm}
\end{figure}
Fig.~\ref{LTR-omega} shows the LTR relic density of neutralinos in the same 1,700 MSSMs (the same whose standard relic density is shown in Fig.~\ref{well-tempered}), each shown as one point in each panel.  Fig.~\ref{LTR-omega} shows that all points can be brought to cross the right DM density  cyan line with suited combinations (in general not unique) of $T_{RH}$ and $\eta= b~100{\rm TeV} / m_\phi$. This means that neutralinos can have the DM density in (almost) all supersymmetric models, provided the right values of $T_{RH}$ and  $\eta$ can be obtained (the exception being severely overabundant or underabundant very light neutralinos, rarely encountered in supersymmetric models~\cite{Gelmini:2006pq}). This has important implications not only for colliders but for direct and indirect DM searches as well. For example, the region of viable supersymmetric models to be searched for in direct detection experiments extends  from  well under 1 GeV to  10 TeV in neutralino mass~\cite{ggsy-2}.
 \begin{figure}
\includegraphics[width=0.40\textwidth]{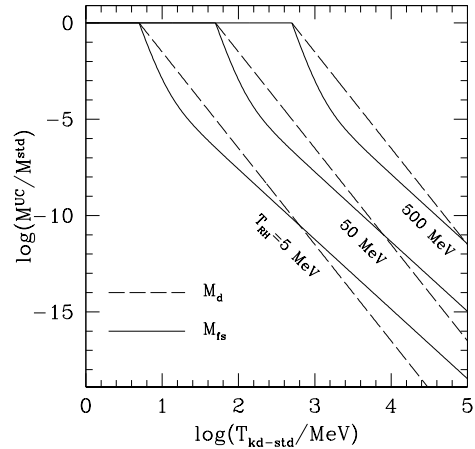}
\vspace{-10pt}
\caption{Mass within the kinetic decoupling horizon and free-streaming volumes ($M_d$ and $M_{fs}$, respectively) of Ultra Cold WIMPs with respect to their standard values as function of the standard kinetic decoupling temperature for different reheating temperatures $T_{RH}$~\cite{Gelmini:2008sh}.
}
\label{UCW}
\vspace{-0.6cm}

\end{figure}

Not only the relic density of WIMPs but their characteristic speed before structure formation in the Universe can differ in standard and non-standard pre-BBN cosmological models.	 
 If kinetic decoupling  happens during the reheating phase of LTR models, WIMPs can have much smaller characteristic speeds, i.e.  be much ``colder" ~\cite{Gelmini:2008sh}, with free-streaming lengths several orders of magnitude smaller than in the standard scenario (see Fig.~\ref{UCW}). Much smaller DM structures  could thus be formed, a fraction of which may persist as clumps within  galactic haloes and be detected in indirect DM searches. The signature would be a much larger boost factor of the annihilation signal than expected in standard cosmologies for a particular WIMP candidate.  WIMPs may instead  be much ``hotter"  than in standard cosmologies too, they may even be warm DM instead of cold, which would leave an imprint on the large scale structure spectrum ~\cite{Lin:2000qq}.

Let us add that also finding ``visible" sterile neutrinos, i.e. those that could be found soon in neutrino experiments, would be a signature of a non standard pre-BBN cosmology. If they are produced through oscillations, they are produced mostly at $T\simeq 13 {\rm\,MeV} (m_s/ 1{\rm\,eV})^{1/3}$, which is $> $4 MeV if  their mass is $m_s > 10^{-3}$ eV . In order to be found in experiments, these sterile neutrinos would necessarily have mixings with active ones large enough to be overabundant, and thus be rejected, in standard cosmologies. In LTR models the relic abundance of visible sterile neutrinos  could be reduced enough for them to be cosmologically acceptable~\cite{visible-sterile}.

 \section{Colliders as DM and pre-BBN cosmology probes}  
 
 In most scenarios one can think of the LHC should find at least  a hint of  new physics, and whatever it finds will lead to a set of possible DM candidates and reject others. In our most ambitious scenarios, the
 LHC will be able to determine a range for the annihilation cross section of the DM candidate, thus a range for the standard relic density $\Omega_\chi^{std}$. For example, the standard relic density  in the LCC2 model could be determined within 40\%, and the error could be reduced to 14\%  with the ILC-500 GeV and to 8\%  with the LHC-1000 GeV~\cite{Baltz:2006fm}.

Assume that the LHC finds a DM candidate and determines its standard relic density  to be  much larger than the DM density. Then, either the particle is unstable (e.g. in SUSY models it is the next-to-lightest particle, the NLSP, and the SUSY spectrum should tell if this is true) or it does constitute the DM (found in DM searches) and the pre-BBN cosmology is non standard.  

The LHC may instead determine that the standard relic density of the DM candidate is much smaller than the DM density. In this case  there are at least two possibilities: either this is just a part of the DM and the DM has other components (which are to be found through DM detection experiments), or this particle constitutes the bulk of the DM (and this will need to be determined through DM searches) and the pre-BBN cosmology is non standard. The separation between these two possibilities will be difficult, in part because the interaction rate in direct DM detection experiments may be large even for very underabundant DM halo particles~\cite{Duda:2001ae}. Indirect DM searches may be very important in this case~\cite{Duda:2002hf}.

If  $\Omega_\chi^{std}$ of a DM candidate is found to be compatible, within the error in the determination, with the DM density, there are still at least two possibilities. One is that the DM candidate found is actually not the DM (namely it is not found in DM searches even if the interaction cross section is large enough). In this case either the DM candidate found is unstable and decays outside the detectors, or the pre-BBN cosmology is non-standard and the particle has a very small relic density.  The second possibility is that we have truly found the main component of the DM.  Even in this case we still would want to get bounds on the possible departure of the pre-BBN cosmology from standard, given that this particle is the only relic we have from this early epoch in the history of the Universe.  This, in fact, may be a good argument (one among many) to build the next collider after the LHC.

 In summary, so far we  have explored how well the LHC and ILC could determine the relic density largely assuming that the pre-BBN evolution of the Universe is standard.  In reality, with the LHC and  ILC we are trying to measure a combination of both, the DM relic density and key parameters of the pre-BBN cosmology which are necessarily tied up.    To disentangle both types of parameters we need to combine accelerator measurements with direct and indirect DM searches, which would tell us the actual halo density of the DM particle candidate. 
 
 The best possible calculation of the WIMP relic density (and relic velocity distributions) assuming a standard cosmology are necessary to eventually discriminate between standard and non standard pre-BBN cosmologies, but not to reject elementary particle models at this time (no ``DM constraint" should be used to reject particle models).
 
 Few attempts have been made so far to figure out how to discriminate particle physics parameters and pre-BBN cosmological parameters at colliders. 
 For example, Ref.~\cite{Chung:2007cn} studied benchmark points LCC1', 2', 3' and 4', slightly shifted  in parameter space with respect to the unprimed LCC points, so that  neutralinos are underabundant in the standard cosmology. Assuming  neutralinos in these models  decoupled while the Universe was in a kinetion period, a simultaneous search could be made at the LHC and ILC for the relic density and for the kinetion  parameter $\eta_\phi$. Ref~\cite{Chung:2007cn} finds that in some of these models the LHC can distinguish that the relic density is the DM density for $\eta_\phi$ non-zero, but in others the LHC finds that the models are compatible with the standard cosmology. In all cases the ILC would do a good job of determining both parameters.

  Much more work is needed  to try to figure out how to discriminate particle physics parameters and pre-BBN cosmological parameters using the data of colliders and other experiments on DM candidates.

\end{document}